\documentclass[12pt,preprint]{aastex}

\slugcomment{Not to appear in Nonlearned J., 45.}

\shorttitle{Evolution of [\ion{O}{3}] Line Profile}
\shortauthors{Wang et al.}

\begin{document}


\title{EVOLUTION OF [\ion{O}{3}]$\lambda$5007 EMISSION-LINE PROFILES IN NARROW EMISSION-LINE GALAXIES}


\author{J. Wang Y. F. Mao and J. Y. Wei}
\affil{National Astronomical Observatories, Chinese Academy of Sciences, Beijing, China, 100021}

\email{wj@bao.ac.cn}




\begin{abstract}

The AGN-host co-evolution issue is investigated here by focusing on the evolution of the 
[\ion{O}{3}]$\lambda5007$ emission-line profile. In order to simultaneously measure both [\ion{O}{3}] line 
profile and circumnuclear stellar population in individual spectrum, a 
large sample of narrow emission-line galaxies is selected from the 
MPA/JHU SDSS DR7 catalog. By requiring that 1) the [\ion{O}{3}] line signal-to-noise ratio
is larger than 30, 2) the [\ion{O}{3}] line width is larger than the instrumental resolution 
by a factor of 2, our sample finally contains 2,333 Seyfert galaxies/LINERs (AGNs),  
793 transition galaxies, and 190 starforming galaxies. 
In additional to the commonly used profile parameters (i.e., line centroid, relative velocity shift and
velocity dispersion), two dimensionless shape parameters, skewness and kurtosis, are used to quantify
the line shape deviation from a pure Gaussian function. We show that the transition galaxies are systematically associated
with narrower line widths and weaker [\ion{O}{3}] broad wings than the AGNs, which
implies that the kinematics of the emission-line gas is different in the two kinds of objects.
By combining the measured host properties and line shape parameters,
we find that the AGNs with stronger blue asymmetries tend to
be associated with younger stellar populations. However, the similar trend is not identified in the 
transition galaxies. The failure is likely resulted from a selection effect in which 
the transition galaxies are systematically associated with younger stellar populations than the AGNs.
The evolutionary significance revealed here suggests that 
both NLR kinematics and outflow feedback in AGNs co-evolve with their host galaxies.
\end{abstract}


\keywords{galaxies: active --- galaxies: evolution --- quasars: emission lines}



\section{Introduction}

The emission from narrow-line region (NLR) of active galactic nucleus (AGN) is an important tool 
to study the relation between the activity of the central supermassive black hole (SMBH)  
and the growth of its host galaxy in which the SMBH resides, both because the NLR emission is 
mainly resulted from the illumination by the central AGN and because the NLR kinematics is 
believed to be mainly dominated by the gravity of the bulge (see review in Wilson \& Heckman 1985 
and references therein, Whittle 1992a,b; Nelson \& White 1996). 
The gravity dominated kinematics motivates 
a number of previous studies to demonstrate that the line width of the AGN's strong 
[\ion{O}{3}]$\lambda$5007 emission line can be used as a proxy for the stellar velocity
dispersion of the bugle (e.g., Nelson \& White 1996; Nelson 2000; Boroson 2003; 
Komossa \& Xu 2007). Basing upon the tight $M_{\mathrm{BH}}-\sigma_*$ relationship
(e.g., Tremaine et al. 2002; Ferrarese \& Merritt 2000; Magorrian et al. 1998; 
Gebhardt et al. 2000; Haring \& Rix 2004), the proxy therefore allows one to 
easily estimate 
$M_{\mathrm{BH}}$ in a large sample of AGNs (e.g., Grupe \& Mathur 2004; 
Wang \& Lu 2001; Komossa \& Xu 2007).

It is well known for a long time that the line profiles of the [\ion{O}{3}] doublelets 
show a blue asymmetry with an extended blue wing and a sharp red falloff in a large fraction 
of AGNs (e.g., Heckman et al. 1981;
Whittle 1985; Wilson \& Heckman 1985; Grupe et al 1999; Tadhunter et al. 2001; 
Veron-Cetty et al. 2001; Zamanov et al. 2002; Komossa \& Xu 2007; Xu \& Komossa 2009; 
Greene \& Ho 2005; de Roberties \& Osterbrock 1984; Storchi-Bergmann et al. 1992;
Arribas et al. 1996; Christopoulou et al. 1997).
The blue asymmetry requires a narrow core Gaussian profile ($\rm FWHM\sim200-500\ km\ s^{-1}$)
with a blueshifted, broad Gaussian component ($\rm FWHM\sim500-1000\ km\ s^{-1}$) to reproduce the
observed asymmetric profiles for both [\ion{O}{3}]$\lambda\lambda4959, 5007$ emission lines.
The spectroscopic monitor revealed a variability time 
scale from one to ten years for the blue wings of the [\ion{O}{3}]$\lambda5007$ lines in two
type I AGNs (IZW\,1: Wang et al. 2005; NGC\,5548: Sergeev et al. 1997),
which means that the blue wings are likely emitted from the 
intermediate-line region located between the traditional BLR and NLR.
In addition to the blue asymmetry, the redshifts of the [\ion{O}{3}] doublelets
are often found to be negative compared to the redshifts measured from both  
stellar absorption features and H$\beta$ emission line 
(i.e., [\ion{O}{3}] blueshifts, e.g., Phillips 1976; Zamanov et al .2002;
Marziani et al. 2003; Aoki et al. 2005; Boroson 2005; Bian et al. 2005; Komossa et al. 2008).
Although they are rare cases, the objects with strong [\ion{O}{3}] blueshifts larger than
100$\mathrm{km\ s^{-1}}$ are called ``blue outliers''.

The popular explanation of the observed [\ion{O}{3}] emission-line profile is that the material outflow 
from central AGN plays important role in reproducing the observed blue asymmetry and blueshift.
With the advent of the high spatial resolution of Hubble Space Telescope (HST), 
spatially resolved spectroscopic observations of a few nearby Seyfert 2 galaxies indicate that the 
NLRs show complicate kinematics, which could reproduce the observed [\ion{O}{3}] line profiles by the 
radial outflow acceleration (or deceleration) and/or jet expansion 
(e.g., Crenshaw et al. 2000; Crenshaw \& Kraemer 2000; Ruiz et al. 2001; Nelson et al. 2000;
Hutchings et al. 1998; Das et al. 2005, 2006, 2007; Kaiser et al. 2000; Crenshaw et al. 2010,
Schlesinger et al. 2009; Fischer et al. 2010; Fischer et al. 2011).

Recent systematical studies suggested that the blue asymmetry is related with
the activity of the central SMBH. Veron-Cetty et al. (2001) indicated that half of their sample of
narrow-line Seyfert 1 galaxies (NLS1s) shows a broad and blueshifted [\ion{O}{3}]$\lambda$5007
component in addition to the unshifted narrow core component. Nelson et al. (2004) found a 
correlation between the blue asymmetry and Eigenvector-I space
by studying the [\ion{O}{3}]$\lambda$5007 line profiles of the PG quasars.
The quasars associated with larger blue asymmetries tend to be stronger \ion{Fe}{2} emitters 
presumably having larger Eddington ratios ($L/L_{\mathrm{Edd}}$, where
$L_{\rm Edd}=1.26\times10^{38}(M_{\rm BH}/M_\odot)\ \mathrm{erg\ s^{-1}}$ is the Eddington 
luminosity, see also in Xu et al. 2007; Boroson 2005; Greene \& Ho 2005; 
Mathur \& Grupe 2005). Similar as the blue asymmetry, the [\ion{O}{3}] blueshift is also 
found to be related with a number of AGN properties. Some authors claimed that the 
[\ion{O}{3}] blueshift is directly correlated with  $L/L_{\mathrm{Edd}}$ (e.g., Boroson 2005; 
Bian et al. 2005), although the correlation might not be the truth (e.g., Aoki et al. 2005).
Marziani et al. (2003) pointed out that all the ``blue outliers'' have small H$\beta$ line widths
($<4000\mathrm{km\ s^{-1}}$) and high $L/L_{\mathrm{Edd}}$ (see also in Zamanov et al. 2002;
Komossa et al. 2008).

AGNs are now widely believed to co-evolve with their host galaxies, which is implied by the tight
$M_{\mathrm{BH}}-\sigma_*$ correlation (see the citations in the first paragraph) and by the 
global evolutionary history of the growth of the central SMBH that traces the star formation history
closely from present to $z\sim5$ (e.g., Nandra et al. 2005; Silverman et al. 2008; 
Shankar et al. 2009; Hasinger et al. 2005). A number of studies recently provided direct evidence 
supporting the co-evolutionary scenario in which an AGN evolves along the Eigenvector-I space 
from a high $L/L_{\mathrm{Edd}}$ state to a low $L/L_{\mathrm{Edd}}$ state as the circumnuclear 
stellar population continually ages (e.g., Wang et al. 2006; 
Wang \& Wei 2008, 2010; Kewley et al. 2006; Wild et al. 2007; Davis et al. 2007).
The results of theoretical simulations indicate a possibility that a major merger between two gas-rich disk galaxies
plays important role in the co-evolution of AGNs and their host galaxies (e.g., Di Matteo et al. 2007;
Hopkins et al. 2007; Granato et al. 2004).
Detailed analysis even suggests a delay of $\sim10^2$Myr for the detectable AGN activity after the 
onset of the star formation activity (e.g., Wang \& Wei 2005; Schawinski et al. 2009; Hopkins et al. 2005; 
Davis et al. 2007; Wild et al. 2010). The delay might be resulted from the feedback from either
the central AGN (e.g., Hopkins et al. 2005) or the stellar winds (e.g., Morman \& Scoville 1988), and/or 
resulted from the purely dynamical origin (Hopkins 2011).

The outflow 
origin for both blue asymmetry and blueshift naturally gives us the hint that the [\ion{O}{3}] emission-line 
profile (and the inferred NLR kinematics) co-evolves with the stellar population of the host galaxy. 
We here report a study that is an effort to examine the evolution of the [\ion{O}{3}]$\lambda$5007 
emission-line profile.  In principle, both broad- and narrow-line AGNs are needed to be analyzed to give a
complete study. Although the [\ion{O}{3}] line profiles can be easily measured in the spectra of typical type I AGNs, the
host galaxy properties are hard to be determined because of the strong contamination from the central AGN's continuum.
Since our aim is to study the relationship between the line profile and the host galaxy properties, 
the narrow emission-line galaxies from Sloan Digital Sky Survey (SDSS) are adopted in our 
study because their spectra allow us to  simultaneously measure both [\ion{O}{3}]$\lambda$5007 line profile and 
stellar population properties in individual object. The high-order dimensionless line
shape parameters are adopted by us to describe the line profile in details. 

The paper is organized as follows. 
The sample selection is presented in \S2. \S3 describes the data reduction, including the 
stellar component removal, the line profile measurements, and the stellar population age measurements. 
Our analysis and results are shown in \S4. The implications are discussed in the next section. A $\Lambda$
cold dark matter ($\Lambda$CDM) cosmology with parameters $H_0=70\ \mathrm{km\ s^{-1}\ Mpc^{-1}}$, 
$\Omega_m=0.3$, and $\Omega_{\Lambda}=0.7$ (Spergle et al. 2003) is adopted throughout the paper.

\section{SAMPLE SELECTION}

\subsection{General selection based on MPA/JHU catalog}

We select a sample of narrow emission-line galaxies from the value-added
SDSS Data Release 7 (Abazajian et al. 2009) MPA/JHU catalog (Heckman et al. 2004; 
Kauffmann et al. 2003a,b,c; Tremonti et al. 2004; see Heckman \& Kauffmann 2006 for a review)
\footnote{These catalogs can be downloaded from http://www.mpa-garching.mpg.de/SDSS/DR7/.},
The catalog provides the measured spectral and physical properties of 927,555 individual narrow emission-line galaxies, 
including Seyfert galaxies, LINERs, transition galaxies and star-forming galaxies.
The integrated spectra of the narrow emission-line galaxies allow us to study the relationship between 
the [\ion{O}{3}]$\lambda$5007 emission-line profiles and the circumnuclear stellar populations, because their 
continuum is dominated by the starlight components.

At first, we select the galaxies whose spectra have median signal-to-noise ratio per pixel of the whole spectrum 
S/N$>$20 from the MPA/JHU SDSS DR7 catalog to ensure that the stellar features could be properly separated 
from each observed spectrum. Secondly, all the emission lines used in the traditional three 
Baldwin-Phillips-Terlevich (BPT) diagnostic 
diagrams\footnote{The BPT diagrams were originally proposed by Baldwin et al. (1981), and then refined by 
Veilleux \& Osterbrock (1987), which are commonly used to determine the dominated powering source in 
narrow emission-line galaxies through their emission-line ratios.} are required to be detected 
with a significant level at least 3$\sigma$. Thirdly, 
the spectral profiles of the [\ion{O}{3}]$\lambda$5007 emission lines are only 
measured for the galaxies whose [\ion{O}{3}] lines have S/N larger than 30. 
Finally, the galaxies with redshifts within the range from 0.11 to 0.12 are removed in the subsequent spectral
analysis. The removement is used to avoid the possible fake spectral profile caused by the poorly 
subtracted strong sky emission line [\ion{O}{1}]$\lambda$5577 at the observer frame. 

\subsection{Instrumental resolution selection}

After the selection based on the MPA/JHU catalog, we remove all the galaxies whose [\ion{O}{3}] line widths
are smaller than $2\sigma_{\mathrm{inst}}$ from our subsequent statistical comparisons according to our spectral profile measurements
(see Section 3 for details). The mean intrinsic resolution used 
in the above criterion is adopted to be 65$\mathrm{km\ s^{-1}}$ according to the SDSS pipelines measurements 
(e.g., York et al. 2000). There are two reasons for this instrumental resolution selection. At first, the cut on line
width removes the low mass starforming galaxies without bugle. Secondly, the cut can alleviate the impact on the 
measured line shape parameters caused by the instrumental resolution.
For a pure Gaussian profile, 
the intrinsic line width $\sigma$ can be obtained by the equation $\sigma^2=\sigma^2_{\mathrm{obs}}-\sigma^2_{\mathrm{inst}}$, where $\sigma_{\mathrm{obs}}$ and $\sigma_{\mathrm{inst}}$ is the observed line width and the instrumental resolution, respectively. 
However, this relationship is only a first-order approximation for the line profiles that 
deviate from a pure Gaussian profile. Our analytic calculation indicates that the correction of the instrumental 
resolution depends on the deviation not only for the line width, but 
also for the other two high-order dimensionless line shape parameters (see Section 4 and Appendix A for details).

\subsection{Final sample}

After removing the duplicates in the MPA/JHU catalog, there are in total 3,339 entries fulfilling the
above selection criteria. The spectra with the highest S/N ratios are kept in our final sample in the 
duplicate removements.
Given the line fluxes reported in the MPA/JHU catalog, these 
galaxies are subsequently classified into three sub-samples according to the 
[\ion{O}{3}]/H$\beta$ versus [\ion{N}{2}]/H$\alpha$ diagnostic diagram using the widely accepted 
classification schemes (Kauffmann et al. 2003a; Kewley et al. 2001). 
Out of the 3,339 galaxies, there are 
2,333 Seyfert galaxies/LINERs (hereafter AGNs for abbreviation), 
793 transition galaxies located between the empirical and the theoretical demarcation lines, and 
190 star-forming galaxies. 

There are 5,856 star-forming galaxies with line width $\sigma<2\sigma_{\mathrm{inst}}$, 
whose line profiles are believed to be dominated by a symmetric Gaussian function.
These galaxies are used as a comparison sample to present the scatters of the measured high-order dimensionless
line shape parameters for a symmetric Gaussian line profile, because of the limited wavelength sampling and 
limited signal-to-noise ratio. These scatters are useful in determining how 
strong a line profile deviates from a symmetric Gaussian profile. 

The three diagnostic BPT diagrams are shown in Figure 1 for the final samples,
in which the AGNs, transition galaxies 
and star-forming galaxies are symbolized by the red, green and blue solid squares, respectively.
The contours represent the distributions of the star-forming galaxies in the comparison sample.





\section{SPECTRAL ANALYSIS}

\subsection{Stellar Features Separation} \label{bozomath}

The spectra of the selected objects are reduced and analyzed by our 
principal component analysis (PCA) pipeline (see Wang \& Wei (2008) for details).
At first, each spectrum is corrected for the Galactic extinction 
using the extinction law with $R_V=3.1$ (Cardelli et al. 1989), in which 
the color excess $E(B-V)$ is taken from the Schlegel, Finkbeiner, and
Davis Galactic reddening map (Schlegel et al. 1998). Secondly, each 
extinctions-corrected spectrum is then transformed to the 
rest frame, along with the flux correction due to the relativity effect, given the redshift 
provided by the SDSS pipelines (Glazebrook et al. 1998; Bromley et al. 1998).
The stellar absorption features are subsequently separated from each rest-frame spectrum
by modeling the continuum and absorption features by the sum of the first seven 
eigenspectra. The eigenspectra are built from the standard single stellar population spectral
library developed by Bruzual \& Charlot (2003). Each of the spectra is fitted
over the rest-frame wavelength range from 3700 to 7000\AA\ by a $\chi^2$ minimization, 
except for the regions with strong emission lines (e.g., Balmer lines, [\ion{O}{3}]$\lambda\lambda$4959, 5007,
[\ion{N}{2}]$\lambda\lambda$6548,6583, [\ion{S}{2}]$\lambda\lambda$6716, 6731, [\ion{O}{2}]$\lambda$3727,
[\ion{O}{3}]$\lambda$4363 and [\ion{O}{1}]$\lambda$6300).

\subsection{Emission-line Profile Measurements}

\subsubsection{Line profile parameters}

The rest-frame emission-line isolated spectra are used to parametrize the spectral profiles of their 
[\ion{O}{3}]$\lambda$5007 emission lines, after the stellar components are 
removed from the observed spectra. The emission-line profile can be parametrized 
by many possible ways. The widely used ones include the FWHM and the second moment 
of the line that is defined as 
\begin{equation}
\sigma^2=\bigg(\frac{c}{\overline{\lambda}}\bigg)^2\frac{\int (\lambda-\overline{\lambda})^2f_\lambda d\lambda}{\int f_\lambda d\lambda}
\end{equation}  
where $\sigma$ is in units of $\rm km\ s^{-1}$, 
$\overline{\lambda}=\int\lambda f_\lambda d\lambda/\int f_\lambda d\lambda$ and $f_\lambda$ is the line 
centroid (i.e., the first moment) and the flux density of the continuum-subtracted line flux, respectively. 
For a pure Gaussian profile, we have a relationship
$\mathrm{FWHM}=2\sqrt{2\ln2}\sigma\approx2.35\sigma$, which means that both FWHM and $\sigma$ comparably
describe the line broadening if the line profile is a Gaussian function. However, as stated in Greene \& Ho (2005), 
$\sigma$ contains more information on the line profile broadening if the profile deviates from a pure Gaussian 
profile.

Heckman et al. (1981) defined an asymmetry index $\rm{AI_{20}=(WL20-WR20)/(WL20+WR20)}$ 
to measure the line asymmetry of [\ion{O}{3}] emission lines, where $\rm WL20$ and $\rm WR20$ is the 
half width to the left and right of the line center defined as the 80\% peak intensity level.  
Basing upon the index $\rm{AI_{20}}$, the authors found that the blue asymmetry is shown in about 80\% of the 36  
Seyfert and radio galaxies. Alternatively, the asymmetry index 
$\rm A.I.= [C(\frac{3}{4})-C(\frac{1}{4})]/FWHM$ is widely used to measure the asymmetry of 
AGN's broad emission lines, where $\rm C(i)$ are profile centroids measured at different levels (see review in 
Sulentic et al. 2000). In addition to the asymmetry, the emission-line shapes of AGNs vary from 
extremely peaked to very ``boxy''. Previous studies used the ratio of line widths at different levels 
to parametrize such line shapes (e.g., Marziani et al. 1996).  

To make the statistical study on [\ion{O}{3}] line profile is feasible for a large sample, 
we adopt the high-order dimensionless 
line shape parameters to describe the line profile departures from a pure Gaussian function 
(see Binney \& Merrifield (1998) for more details): 
\begin{equation}
\xi_k=\mu_k/\sigma^k\ k\geq3
\end{equation} 
where $\mu_k$ is the $k$-order moment defined as 
\begin{equation}
\mu_k=\bigg(\frac{c}{\overline{\lambda}}\bigg)^k\int (\lambda-\overline{\lambda})^kf_\lambda d\lambda
\end{equation}
and $\sigma$ is the second order moment defined in Eq (1), respectively. 
The first shape parameter $\xi_3$ is termed the ``skewness'' that measures the deviation from symmetry.
A pure Gaussian profile corresponds to $\xi_3=0$. The emission-line shape with a positive value of $\xi_3$ 
shows a red asymmetry, and the shape with a negative value a blue asymmetry. The second parameter 
$\xi_4$ is termed the ``kurtosis'' that measures the symmetric deviation from a pure Gaussian profile 
(for a pure Gaussian profile, we have $\xi_4=3$).  The emission line with peaked profile superposed on a 
broad base has a value of $\xi_4>3$.
A value of $\xi_4<3$ corresponds to a ``boxy'' line profile. We refer the readers to Figure 11.5 in 
Binney \& Merrifield (1998) for how the values of $\xi_3$ and $\xi_4$ vary with the line shapes.

\subsubsection{Relative velocity shifts}

The bulk relative velocity shift of the [\ion{O}{3}] emission line is measured relative to the H$\beta$ emission line
in each emission-line isolated spectrum, because the H$\beta$ line can be easily detected and because the line 
shows very small velocity shift relative to the galaxy rest-frame (e.g., Komossa et al. 2008). The 
velocity shift is calculated as $\Delta\upsilon=\delta\lambda/\lambda_0 c$, where 
$\lambda_0$ and $c$ is the rest-frame wavelength of the
[\ion{O}{3}]$\lambda5007$ emission line and the velocity of light, respectively. $\delta\lambda$ denotes
the wavelength shift of the [\ion{O}{3}] line with respect to the H$\beta$ line. The shift is determined 
from the measured line centroids ($\overline{\lambda}$) and the rest-frame wavelengths in vacuum of both lines. 
With the definition of the
velocity shift, a blueshift corresponds to a negative value of $\Delta\upsilon$, and a redshift to a 
positive value.

\subsection{Stellar Population Properties}
 
By following our previous studies again, we use the two Lick indices, i.e., the 4000\AA\ break ($D_n(4000)$)
and the equivalent width of H$\delta$ absorption feature of A-type stars (H$\delta_A$)\footnote{
The 4000\AA\ break is defined as 
$D_n(4000)=\int_{4000}^{4100}f_\lambda d\lambda/\int_{3850}^{3950}f_\lambda d\lambda$
(Bruzual 1983; Balogh et al. 1999). The index H$\delta_{\rm{A}}$ is defined
by Worthey \& Ottaviani (1997) as
$\mathrm{H}\delta_A=(4122.25-4083.50)(1-F_I/F_C)$
where $F_I$ is the flux within the $\lambda\lambda4083.50-4122.25$ feature bandpass, and $F_C$ the flux of
the pseudo-continuum within two defined bandpasses: blue $\lambda\lambda4041.60-4079.75$ and
red $\lambda\lambda4128.50-4161.00$.}, as 
the indicators of the ages of the stellar populations of the galaxies (e.g., Heckman et al. 2004; 
Kauffmann et al. 2003; Kauffmann \& Heckman 2008; Kewley et al. 2006; Wang \& Wei 2008, 2010; Wild et al. 2007). 
Both indices are measured in the removed stellar component for each spectrum, and are
reliable age indicators until a few Gyr after the onset of a  burst (e.g., Kauffmann et al. 2003, Bruzual \& Charlot 2003).

\subsection{Broad-Line AGNs}

The sub-sample of partially obscured AGNs associated with broad H$\alpha$ emission lines 
(i.e., Seyfert 1.8/1.9 galaxies) is selected from the parent AGN and transition galaxy samples.
Although the existence of the broad H$\alpha$ emission is direct indicative of the accretion activity of the
central SMBH, the underlying AGN's continuum in these partially obscured AGNs could potentially decrease the
measured two indices. These partially obscured AGNs are therefore removed from the 
subsequent analysis when the two indices are involved.

Following our previous studies (Wang \& Wei 2008), 
the broad-line AGNs are selected by the means of the high-velocity wings of the broad 
H$\alpha$ components on their blue side after the stellar features are removed from 
the spectra. The red wing is not adopted in the selection because of the superposition of 
the strong [\ion{N}{2}]$\lambda$6583 emission line. The partially obscured AGNs are at first 
automatically selected by the criterion $F_{\mathrm{w}}/\sigma_c\geq3$, where $F_{\mathrm{w}}$ 
is the specific flux of the 
line wing averaged within the wavelength range from 6500 to 6350\AA\ in the rest-frame, 
and $\sigma_c$ is the standard deviation of the continuum flux within the emission-line
free region ranging from $\lambda5980$ and $\lambda6020$. The automatically selected   
AGNs are then inspected one by one by eyes. In total, there are 174 broad-line Seyfert
galaxies/LINERs and 55 broad-line transition galaxies.

\subsection{Uncertainties estimation}

The MPA/JHU catalog contains many duplicates because of the repeat observations. The duplicates 
allow us to roughly estimate the uncertainties for the line shape parameters and the stellar population 
age indicators. The duplicates are reduced and measured by the same method as described above. For the three main
sub-samples (i.e., AGNs, transition galaxies and star-forming galaxies), the uncertainties are estimated to 
be $0.14\pm0.08$ and $0.20\pm0.14$ for the shape parameters $\xi_3$ and $\xi_4$, respectively. The two Lick indices
have uncertainties: $\Delta D_n(4000)=0.03\pm0.02$, and $\Delta$H$\delta_A=0.38\pm0.19$. 
For the comparison sample, the duplicates provide the uncertainties of $0.08\pm0.05$ and $0.13\pm0.06$ for the 
parameters $\xi_3$ and $\xi_4$, respectively.






\section{ANALYSIS AND RESULTS}

After obtaining all the required parameters, the relation between the [\ion{O}{3}] line 
profiles and the stellar population properties is examined in this section.

\subsection{[\ion{O}{3}] Line Width and [\ion{O}{3}]$\lambda5007$ Relative Velocity Shifts: AGNs versus Transition Galaxies}

The [\ion{O}{3}] line profile is compared between the AGNs and the transition galaxies in this section.
To avoid the possible systematics caused by the host galaxy properties, we compare the line profiles
in a sample of matched galaxy pairs (e.g., Kauffmann et al. 2006). The matched pairs are created according to the 
SDSS DR4 MPA/JHU AGN catalog. The catalog provides the measurements of galaxy properties for the AGNs that are 
classified according to the demarcation line given in Kauffmann et al. (2003). We select the pairs that have
$\Delta\log M_*<0.1$, $\Delta\log\mu_*<0.1$, $\Delta C<0.1$, $\Delta \sigma_*<30\ \mathrm{km\ s^{-1}}$, and 
$\Delta z<0.01$, where $M_*$ is the stellar mass in unit of $M_\odot$, $\mu_*$ is the effective stellar mass-density
($\mu_*=M_*/2\pi r^2_{\mathrm{50,z}}$, where $r_{\mathrm{50,z}}$ is the half-light radius in the $z$-band) in unit 
of $M_\odot\ \mathrm{kpc^{-2}}$, $C$ is the concentration index defined as $C=R_{90}/R_{50}$ the ratio of the radius 
enclosing 90\% of the total flux to that enclosing 50\% of the flux in the $r$-band, $\sigma_*$ is the stellar 
velocity dispersion and $z$ is the redshift. Because there are much more AGNs than the transition galaxies in our sample, a large fraction
of the transition galaxies have a few of AGN pairs. In these cases, one AGN is assigned to a
pair by a random selection, and the final distributions are built from 100 Monte-Carlo iterations. 
Finally, there are totally 163 distinct galaxy pairs, in which no galaxy is assigned to a pair
more than once.

The left panel in Figure 2 compares the cumulative distributions of the [\ion{O}{3}] line widths in terms of the calculated velocity dispersions.
The AGNs and transition galaxies are shown by the red-solid and green-dotted lines, respectively. The error-bars are
resulted from our iterations. 
One can see clearly that
the line profiles of the AGNs are systematically wider than that of the 
transition galaxies. A two-side Kolmogorov Smirnov test yields a maximum discrepancy of 0.282
with a corresponding probability that the two samples match of $4.6\times10^{-5}$.
In fact, the comparison of the parameter $\xi_4$ indicates 
that the difference in the distributions is likely due to the fact that 
the transition galaxies have systematically smaller values of $\xi_4$ than the AGNs 
(see section 4.3 for more details). That means the 
transition galaxies are associated with weaker [\ion{O}{3}] broad components compared to the AGNs,
which implies that the feedback caused by the material outflow is less stronger in the transition galaxies 
than in the AGNs.

The distributions of the [\ion{O}{3}]$\lambda5007$ relative velocity shifts defined in Section 3.2.2 are compared in 
the right panel of Figure 2 between the AGNs and transition galaxies. The symbols are the same as the left panel. 
A two-side Kolmogorov Smirnov test shows that the two samples are matched at a probability of 77\%, 
although a careful examination suggests that the AGNs marginally tend to be associated with larger negative velocity
shifts compared with the transition galaxies.


\subsection{Line Shape Parameters $\xi_3$ and $\xi_4$}

The left-bottom panel in Figure 3 shows the $\xi_4$ versus $\xi_3$ diagram.
The AGNs, transition galaxies and star-forming galaxies are 
symbolized by the red, green and blue points, respectively. The over-plotted contours 
present the distribution of the star-forming galaxies in the comparison sample.
The two crosses at the left-bottom corner show the typical uncertainties for both parameters 
as roughly estimated from the duplicates. The red-solid one indicates the uncertainties for the large
line-width objects, and the black-dashed one for the comparison sample.
The left-upper panel and the right-bottom 
panel shows the distributions of the parameters $\xi_3$ and $\xi_4$ for all the sub-samples 
(including the comparison sample),
respectively. The distributions are color-codded for the sub-samples as same as the  
left-bottom panel. As expected from an emission-line profile dominated 
by the instrumental resolution, the distribution of the comparison sample is strongly clustered 
around the point (i.e., $\xi_3=0$ and $\xi_4=3$) associated with a pure Gaussian function. 
The last column in Table 1 lists the average and median values of $\xi_3$ ($\overline{\xi}_3=0.01$
and $\langle\xi_3\rangle=0.01$) for the comparison sample. The same values are listed in the 
last column in Table 2 but for the parameter $\xi_4$ ($\overline{\xi}_4=2.61$ and $\langle\xi_4\rangle=2.58$).
The dispersions is calculated to be 0.14 and 0.43 for the parameters $\xi_3$ and $\xi_4$, respectively.
The distribution therefore validates our spectral measurements 
because the distribution of the comparison sample is highly consistent with the prediction 
of a pure Gaussian function.

Similar as the comparison sample, the statistics are tabulated in Table 2 and 3 for both
AGNs and transition galaxies as well. One can find from the tables that the line profiles of these 
galaxies systematically deviate from that of the control sample (i.e., deviate from 
a pure Gaussian profile) by not only smaller $\xi_3$ (i.e., a stronger blue asymmetry), but also larger $\xi_4$ 
(i.e., a stronger broad base). The main panel in Figure 3 shows that both AGNs and transition galaxies form 
a sequence starting from the pure Gaussian region to the upper-left corner. 
In fact, this phenomenon is qualitatively in agreement with the fact that 
two Gaussian components, one representing the narrow line core and the other representing  
the blueshifted broad wing, are usually required to model the observed [\ion{O}{3}] line
profiles. A minor fraction of the objects deviate from the sequence by their positive values of $\xi_3$. 
By inspecting the spectra of these objects one by one by eyes, we find that the deviations are mainly resulted from 
the contamination at the [\ion{O}{3}] red wing caused by the weak low-ionization emission line
\ion{He}{1}$\lambda$5016 (Veron et al. 2002).

It is interesting to see that the star-forming galaxies with $\sigma>2\sigma_{\rm inst}$ follow the 
sequence as well, although most of them have Gaussian line profiles. In order to validate the deviations
in these galaxies,
the spectra with $\xi_3<-0.5$ are then inspected one by one by eyes again. 
To understand the nature of these star-forming galaxies with 
blue asymmetric [\ion{O}{3}] emission lines, we need to carefully model all the strong emission lines in 
individual spectrum. The issue is out of the topic of this paper, and will be examined in our 
subsequent studies.

The results from the two-sides Kolmogorov-Smirnov tests are tabulated in Table 3 and 4 for the 
parameters $\xi_3$ and $\xi_4$, respectively. Each entry contains the maximum absolute discrepancy
between the used two sub-samples, and the corresponding probability that the two sub-samples match.  
Although the AGNs and the transition galaxies show similar distributions for $\xi_3$, the AGNs significantly 
differ from the transition galaxies in the distribution of parameter $\xi_4$. 
The fraction of objects with strong broad [\ion{O}{3}] wings is larger in the AGNs than in the transition galaxies

Previous studies frequently identified a strong correlation between the [\ion{O}{3}] line widths 
and the [\ion{O}{3}] blueshifts (e.g., Bian et al. 2005; Komossa et al. 2008; Zamanov et al. 2002). 
The correlation is commonly explained by the material outflows.
The similar correlations are also identified in iron coronal lines (Erkens et al. 1997) and 
optical \ion{Fe}{2} complex (Hu et al. 2008) in typical type I AGNs.  The left panel in Figure 4 plots
the velocity shift as a function of the line width for the AGNs (red points) and the
transition galaxies (green points). Spearman rank-order tests show a marginal anti-correlation 
between the velocity shifts and the line widths for the AGNs
(with a Spearman correlation coefficient $r_s=-0.088$ at a significance level $P<10^{-4}$, where $P$ is the
probability that there is no correlation between the two variables.), and, however, a stronger correlation 
for the transition galaxies (with $r_s=-0.229$ and $P<10^{-4}$). As an additional examination, the 
middle panel in the figure plots the velocity shift against the parameter $\xi_3$.
Strong correlations between the two variables can be identified for both AGNs ($r_s=0.334$ and 
$P<10^{-4}$) and transition galaxies ($r_s=0.513$ and $P<10^{-4}$), which means that larger the blueshift, 
stronger the blue asymmetry. One can see from the correlations that the objects with the largest 
velocity shifts deviate from the correlations. These objects instead show symmetrical line profiles. 
In order to further describe their line profiles, the right panel in the figure plots the velocity shift against 
the parameter $\xi_4$. We find that these [\ion{O}{3}] lines have ``boxy'' line profiles rather 
than peaked ones. In fact, a ``boxy'' line profile could be reproduced by the sum
of two or more (distinct) peaks with comparable fluxes and line widths (see examples of spectra 
in the Figure 1 in Greene \& Ho 2005).

\subsection{[\ion{O}{3}] Line Profile versus Stellar Population}

The evolution effect on the [\ion{O}{3}] line profile is examined in this section by using the 
two indices, $D_n(4000)$ and H$\delta_\mathrm{A}$, as the indicators of the ages of the 
circumnuclear stellar populations. As described in Section 3.3, the partially obscured AGNs are
excluded from the analysis throughout this section.

The two indices are plotted against the parameter
$\xi_3$ (the left panels) and against the relative velocity shift (the right panels) for the
AGNs in Figure 5. 
At the first glance of the four panels, we fail to directly identify any significant 
correlation between the line profile parameters and the stellar population ages. To examine the 
evolution effect in more details, we divide the AGN sample into two groups according to 
their $\xi_3$ values: one group for the galaxies with $\xi_3>-0.5$ and the another one for the 
galaxies with $\xi_3<-0.5$. The distributions of the $D_n(4000)$ values are 
compared between the two groups in the upper-left panel of Figure 7. The comparison indicates a significant 
difference between the two groups\footnote{Note that the similar results can be obtained for the index H$\delta_A$.}. 
The AGNs with larger amount of blue asymmetry are systematically 
associated with younger stellar populations. A two-sides Kolmogorov Smirnov test indicates that 
the difference between the two distributions is at a significance level $P<1\times10^{-9}$ with a maximum absolute 
discrepancy of 0.22. Similar as the parameter $\xi_3$, the AGN sample is instead separated into two groups by the 
relative velocity shift at $\Delta\upsilon=-100\ \mathrm{km\ s^{-1}}$. We identify a marginal trend that larger the blue 
velocity shifts, younger the associated stellar populations (see the bottom-left panel in Figure 7). 
The same statistical test yields a probability 
that the two groups match of $P=9\times10^{-2}$ (with a maximum absolute discrepancy of 0.11).

Figure 6 is the same as Figure 5 but for the transition galaxies. The same methods as the AGN sample are 
adopted to divide these galaxies into two groups. The right panels in Figure 7 compare the 
distributions of $D_n(4000)$ between the two groups. The two-sides 
Kolmogorov-Smirnov tests suggest that the two distributions match at the probabilities $P=0.02$ for 
the $\xi_3$ separation and $P=0.32$ for the $\Delta\upsilon$ separation, which is much less significant
than the AGN sample. In fact, these results are easily understood 
because the transition galaxies are usually systematically younger than the AGNs (see also in Kewley et al. 2006; 
Schawinski et al. 2007; Wang \& Wei 2008).

In summary, our results indicate that the AGNs associated with young stellar populations show a 
wide range in their [\ion{O}{3}] line profiles that vary from a
blue asymmetrical shape to a Gaussian function. In contrast, the
AGNs associated with old stellar populations always show symmetrical line profiles.

\section{DISCUSSIONS}

\subsection{$L/L_{\mathrm{Edd}}$ as A Physical Driver}

The evolution of [\ion{O}{3}]$\lambda$5007 emission-line profile is studied by using a 
large sample of narrow emission-line galaxies selected from the MPA/JHU SDSS DR7 catalog. The  
line-profile parameterizing allows us to reveal a trend that AGNs with more significant blue
asymmetries tend to be associated with younger stellar populations. At the beginning of discussion, we argue that
the trend is not driven only by the orientation effect (Antonucci 1993; Elitzur 2007). It is widely accepted that the 
commonly observed blue asymmetrical and blue-shifted [\ion{O}{3}] line profiles are
caused by the wind-driven NLR outflows\footnote{We refer the readers to the review in Veilleux et al. (2005) 
for the evidence for outflows in AGNs and see Komossa et al. (2008) for a discussion of the various models.}.
The orientation-driven scenario therefore results in
an odd corollary that a fraction of the AGNs with young stellar populations have outflow directions
closer to the line-of-sight of observers than the 
AGNs with old stellar populations, while the starlight components from the host galaxies are believed to be isotropic.

The revealed trend can be alternatively driven by the activity of the central SMBH. In fact,
there is many observational evidence supporting that the [\ion{O}{3}] line blue asymmetries and
blueshifts are correlated with $L/L_{\mathrm{Edd}}$ or Eigenvector-I space\footnote{It is commonly
believed that the Eigenvector-I space is physically driven by $L/L_{\rm Edd}$
(e.g., Boroson \& Green 1992; Boroson 2002).} in typical type I AGNs (see the description and citations
in Section 1). 
The relationship between the [\ion{O}{3}] line profile and $L/L_{\mathrm{Edd}}$ is examined in the current 
narrow-line AGN sample to present a complete understanding. Similar as the previous studies of narrow-line AGNs,
the parameter $L_{\mathrm{[OIII]}}/\sigma_*^4$ (e.g., Heckman et al. 2004) is used as a proxy of 
$\lambda_{\mathrm{Edd}}=L/L_{\mathrm{Edd}}$, 
where $L_{\mathrm{[OIII]}}$ and $\sigma_*$ is the [\ion{O}{3}]$\lambda5007$ 
line luminosity and the bulge velocity dispersion, respectively. The transition galaxies are excluded from the 
subsequent calculations because of the possible contamination caused by \ion{H}{2} regions. The bolometric luminosity
$L$ is transformed from $L_{\mathrm{[OIII]}}$ through the bolometric correction $L/L_{\mathrm{[OIII]}}\approx3500$ 
(see Heckman et al. (2004) for details). By assuming the 
Balmer decrement for the standard Case B recombination and the Galactic extinction curve with $R_V=3.1$,
the [\ion{O}{3}] line luminosity is corrected for the local extinction that is inferred from the narrow-line ratio $\mathrm{H\alpha/H\beta}$. 
The blackhole mass is estimated from the $M_{\mathrm{BH}}-\sigma_*$ calibration: 
$\log(M_{\mathrm{BH}}/M_\odot)=8.13+4.02\log(\sigma_*/200\ \mathrm{km\ s^{-1}})$ (Tremaine et al. 2002), 
in which the velocity dispersion is measured for each spectrum through our PCA fittings. 
The galaxies with $\sigma_*<70\ \mathrm{km\ s^{-1}}$ are removed from the analysis because the SDSS 
instrumental resolution is $\sigma_{\mathrm{inst}}\approx65\ \mathrm{km\ s^{-1}}$. 

The results are shown in Figure 8. The left panel presents an anti-correlation between $L_{\mathrm{[OIII]}}/\sigma_*^4$
and $D_n(4000)$. Larger the Eddington ratio, and younger the stellar population, which is consistent with the 
previous studies (e.g., Kewley et al. 2006; Kauffmann et al. 2007). 
The middle panel plots $L_{\mathrm{[OIII]}}/\sigma_*^4$
as a function of the parameter $\xi_3$.
When compared with Figure 5, one can find a similar trend in the $L_{\mathrm{[OIII]}}/\sigma_*^4$ vs. $\xi_3$ plot
as that in the $D_n(4000)$ vs. $\xi_3$ plane. AGNs with large amount of blue asymmetry tend to
have large $L/L_{\mathrm{Edd}}$. A much stronger trend can be identified in the right panel between
$L_{\mathrm{[OIII]}}/\sigma_*^4$ and the parameter $\xi_4$. The trend indicates a fact that AGNs with
larger Eddington ratios tend to be associated with stronger [\ion{O}{3}] broad components. However, 
we fail to find a significant trend between $L_{\mathrm{[OIII]}}/\sigma_*^4$ and the bulk relative velocity shifts 
$\Delta\upsilon$.

Spearman rank-order tests are performed to show either the Eddington ratio or the stellar population is
intrinsically related with the line asymmetry. The resulted correlation coefficient matrix is listed in Table 5.
All the entries in the table have a probability of null correlation $P<10^{-4}$.
Although both $\lambda_{\mathrm{Edd}}$ and $D_n(4000)$ are correlated with $\xi_3$ at comparable significance levels, 
$\xi_4$ is more strongly correlated with $\lambda_{\mathrm{Edd}}$ than with $D_n(4000)$. 
The stronger correlation therefore indicates that the trend between
the line asymmetry and $D_n(4000)$ is likely driven physically by the evolution of the Eddington ratio (see below). 


\subsection{Evolution of [\ion{O}{3}] Emission-line Profile}

The trend that AGNs with more significant blue asymmetry tend to be associated with younger 
stellar populations provides a piece of direct evidence supporting the co-evolution of the NLR kinematics
and the growth of the host galaxy. AGNs with stronger outflows tend to be at 
their earlier evolutionary stage as inferred from the stellar population ages. The analysis in the above section further suggests that the trend
is likely driven by $L/L_{\mathrm{Edd}}$. The important evolutionary role of $L/L_{\mathrm{Edd}}$ has been frequently proposed in recent studies (e.g., Wang \& Wei 2008, 2010; Kewley et al. 2006; Heckman et al. 2004).
Putting these pieces together yields an improved co-evolution scenario in which AGN likely evolve from a high-$L/L_{\rm Edd}$ state with strong outflow to a low-$L/L_{\rm Edd}$ state with weak outflow as 
the circumnuclear stellar population continually ages.

This evolutionary scenario is consistent with the current understandings of
AGNs. Lieghly et al. (1997) reported evidence of relativity outflows in three NLS1s.
In comparison with typical broad-line Seyfert galaxies, 
NLS1s are ``young'' AGNs (Mathur 2000) typical of smaller $M_{\mathrm{BH}}$, higher $L/L_{\rm Edd}$, larger [\ion{O}{3}] emission-line
asymmetries, younger circumnuclear stellar populations possibly at post-starburst phase, and enhanced star formations as recently revealed by the observation of \it Spitzer\rm\ 
(e.g., Boroson 2005; Zamanov et al. 2002; Boroson \& Green 1992; Wang et al. 1996; Boller et al. 1996; 
Sulentic et al. 2000; Boroson 2002; Xu et al. 2007; Zhou et al. 2005; Wang \& Wei 2006).  
In addition to the observational ground, the feedback from central AGN is frequently involved in the modern numerical 
and sim-analytical galaxy evolution models to quench the star formation and to blow the gas or dust away through the powerful AGN's wind 
(e.g., Springel et al. 2005; di Matteo et al. 2005; Kauffmann \& Heckman 2008; Fabian 1999; Hopkins et al. 2005, 2008a,b; Croton et al. 2006; Somerville et al. 2008; Khalatyan et al. 2008).
The numerical simulation done by Hopkins et al. (2005) suggests a scenario that the 
powerful AGN's wind is required to occur in the young AGN phase in which the central SMBH is heavily obscured by the 
starforming gas and dust.

\subsection{Are Transition Galaxies at Intermediate Evolutionary Phase?}

Transition galaxies are the narrow emission-line galaxies located between the theoretical 
and empirical demarcation lines in the [\ion{O}{3}]/H$\beta$ versus [\ion{N}{2}]/H$\alpha$ diagnostic diagram 
(Kauffmann et al. 2003; Kewley et al. 2001). Their emission-line properties are explained by the mixture of 
the contributions from both star formations and typical AGNs (e.g., Kewley et al. 2006; Ho et al. 1993, see 
recent review in Ho 2008).  
Our spectral profile analysis indicates that the transition galaxies differs from the Seyfert galaxies 
in their [\ion{O}{3}] emission-line profiles. The transition galaxies systematically show weaker [\ion{O}{3}] broad wings, and 
narrower [\ion{O}{3}] line widths than the Seyfert galaxies, which implies that the two kinds of objects are 
different from each other in their mass outflow kinematics.  The line profile comparison allows 
us to suspect that the transition galaxies 
``bridge'' the AGN-starburst co-evolution from early starburst-dominated 
phase to late AGN-dominated phase (e.g., Yuan et al. 2009; Wang \& Wei 2006; Schawinski et al. 2007, 2009). 
The evolutionary role of the 
transition galaxies has been reported in the recent studies by many authors. With the large SDSS spectroscopic 
database, transition galaxies are found to be systematically associated with younger stellar populations than 
Seyfert galaxies (e.g., Kewley et al. 2006; Wild et al. 2007; and also see Figure 6 in this paper).   
Westoby et al. (2007) found that the distribution of the H$\alpha$ equivalent widths of the transition galaxies
peaks at the valley between the red and blue sequences. By analyzing the mm-wavelength observation taken by
the 30m IRAM, Schawinski et al. (2009) suggested that molecular gas reservoir is destructed by the AGN feedback 
at early AGN+starforming transition phase.

\section{Conclusion}  

We systematically examined the evolutionary issue of the [\ion{O}{3}]$\lambda5007$ emission-line profile
by using a large sample of narrow emission-line galaxies selected from the 
MPA/JHU SDSS DR7 value-added catalog. The sample is separated into three sub-samples
(i.e., star-forming galaxies, transition galaxies and Seyfert galaxies/LINERs) basing upon the line ratios given in
the catalog. Two shape parameters, skewness ($\xi_3$) and kurtosis ($\xi_4$), are additionally 
used to quantify the profile deviation from a pure Gaussian. 
Our analysis indicates that a) the transition galaxies are systematically associated
with narrower line widths and with weaker [\ion{O}{3}] broad wings than the AGNs;
b) the AGNs with stronger blue asymmetries tend to
be associated with younger stellar populations. The evolutionary significance of the [\ion{O}{3}] line profile
suggests a co-evolution of the outflow feedback and the AGN's host galaxy.




\acknowledgments
We would like to thank the anonymous referees for his/her constructive comments in 
improving the paper. 
This study was supported by the National Science Foundation of China (under grant
10803008), and by the National Basic Research Program of China (grant 2009CB824800).
This study uses the SDSS archive data that was created and distributed by the
Alfred P. Sloan Foundation.






\appendix

\section{Appendix}

A toy model of emission-line profile is constructed here to illustrate the effect of the 
instrumental resolution on the measured line shape parameters.
The model is constructed by assuming an observed [\ion{O}{3}] emission line profile that is 
prescribed by the sum of two pure Gaussian profiles: $\phi(\upsilon)=\phi_1(0,\sigma_1)+\alpha\phi_2(\mu,\sigma_2)$,
where $\phi_1(0,\sigma_1)$ is a normalized Gaussian distribution with the peak at zero and the 
variance of $\sigma^2_1$, $\phi_2(\mu,\sigma_2)$ is the same distribution but with the peak shifted at $\mu$ and the
variance of $\sigma^2_2$, and $\alpha$ is a parameter much less than unit.  With this two Gaussian components model,
$\phi_1(0,\sigma_1)$ represents the narrow core of the emission, and $\alpha\phi_2(\mu,\sigma_2)$ the shifted, low-contrast
broad wing (We require $|\mu|/\sigma_i\ll 1$ for avoiding double peaked profile). 
The normalizations of the two distributions yield $\int\phi(\upsilon)d\upsilon=1+\alpha$.

The first moment (i.e., the line centroid) is written as 
$\overline{\mu}=\int\upsilon\phi d\upsilon=\int\upsilon\phi_1 d\upsilon+\alpha\int\upsilon\phi_2 d\upsilon=\alpha\mu$,
which is not affected by the instrumental resolution. With the first moment, the high-order moments are written as 
\begin{equation}
E\lbrace(\upsilon-\overline{\mu})^p\rbrace=\int(\upsilon-\overline{\mu})^p\phi d\upsilon 
\end{equation}
The second moment of the total line profile is therefore inferred to be approximately the linear combination of the 
square of the second moments of 
the two distributions: $\int(\upsilon-\overline{\mu})^2\phi d\upsilon=
\sigma^2_1+\alpha\sigma^2_2+\mu^2\alpha(1-\alpha)^2\approx\sigma^2_1+\alpha\sigma^2_2$. In the context of the toy model,
the linear combination means that the instrumental resolution $\sigma_{\mathrm{inst}}$ can be removed from the measured 
values 
according to the equation $\sigma^2_{\mathrm{obs}}=\sigma^2+(1+\alpha)\sigma^2_{\mathrm{inst}}$, 
which means that the removal depends on the relative strength of the low-contrast Gaussian profile, i.e., the parameter $\alpha$.

The third moment of the total profile can be trivially obtained through the integration
$\int(\upsilon-\overline{\mu})^3\phi d\upsilon=(\alpha-3\alpha^2+2\alpha^3-\alpha^4)\mu^3+3\mu\alpha(\sigma^2_2-
\sigma^2_1-\alpha\sigma^2_2)$.
By ignoring the terms with high-orders of $\alpha$, the third moment approximately equals to 
$\approx\alpha\mu^3+3\alpha\mu(\sigma^2_2-\sigma_1^2)$.
The result suggests that the measured third moment depends not only on the relative shift of the 
low-contrast Gaussian profile, but also on the relative strength denoted by the parameter $\alpha$.
Although the second term in the above result means that the third moment does not depend on the 
instrumental resolution in our toy model at the first order approximation: $\sigma_2^2-\sigma_1^2=
\sigma_{\mathrm{obs,2}}^2-\sigma_{\mathrm{obs,1}}^2$, the definition of the dimensionless 
shape parameter $\xi_3$ introduces a non-linear dependence on the instrumental resolution in 
the denominator (see Eq. 2 in the main text). 
By adopting the same approach, the fourth moment is calculated to be 
$\int(\upsilon-\overline{\mu})^4\phi d\upsilon\approx
\alpha\mu^4+3(\sigma^4_1+\alpha\sigma^4_2)+6\alpha\mu^2\sigma^2_2$, where one sees that the 
second and third terms introduce again a non-linear dependence on the instrumental resolution.

With this toy model, the galaxies whose [\ion{O}{3}]$\lambda$5007 lines are narrower
than 2$\sigma_{\mathrm{inst}}$ are then dropped out from our statistical analysis 
in order to avoid the effect of the instrumental resolution on the measured line shape 
parameters.




\clearpage



\begin{figure}
\epsscale{.80}
\plotone{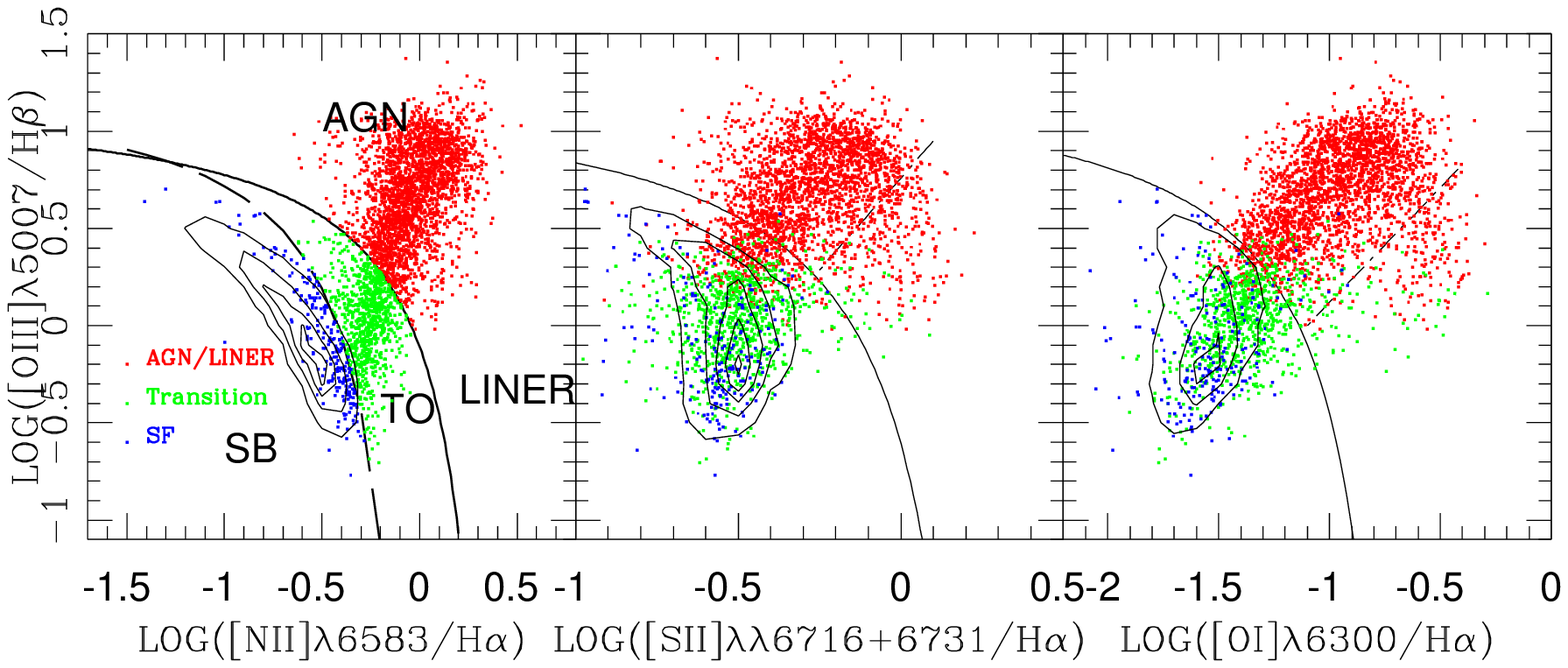}
\caption{Three BPT diagnostic diagrams for the sub-samples. The AGNs, transition galaxies and star-forming galaxies with line widths
$\sigma>2\sigma_{\mathrm{inst}}$ are plotted by the red, green and blue squares, respectively.
The underlying density contours show the distributions on the three diagnostic diagrams for the star-forming galaxies
with narrower emission-line widths, i.e., $\sigma<2\sigma_{\mathrm{inst}}$
The solid lines show the theoretical demarcation lines separating
AGNs from star-forming galaxies proposed by Kewley et al. (2001), and the long-dashed line the empirical line proposed
in Kauffmann et al. (2003).
}
\end{figure}

\begin{figure}
\epsscale{.50}
\plotone{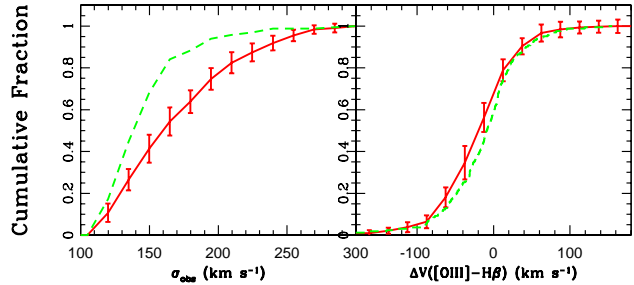}
\caption{\it Left panel:\rm\ Cumulative distributions of the measured line widths of the [\ion{O}{3}] emission 
lines for the AGNs (the red-solid line) and the transition galaxies (the green-dotted line). 
\it Right panel:\rm\ The same as the left panel but for the measured relative velocity shifts
}
\end{figure}

\begin{figure}
\epsscale{.80}
\plotone{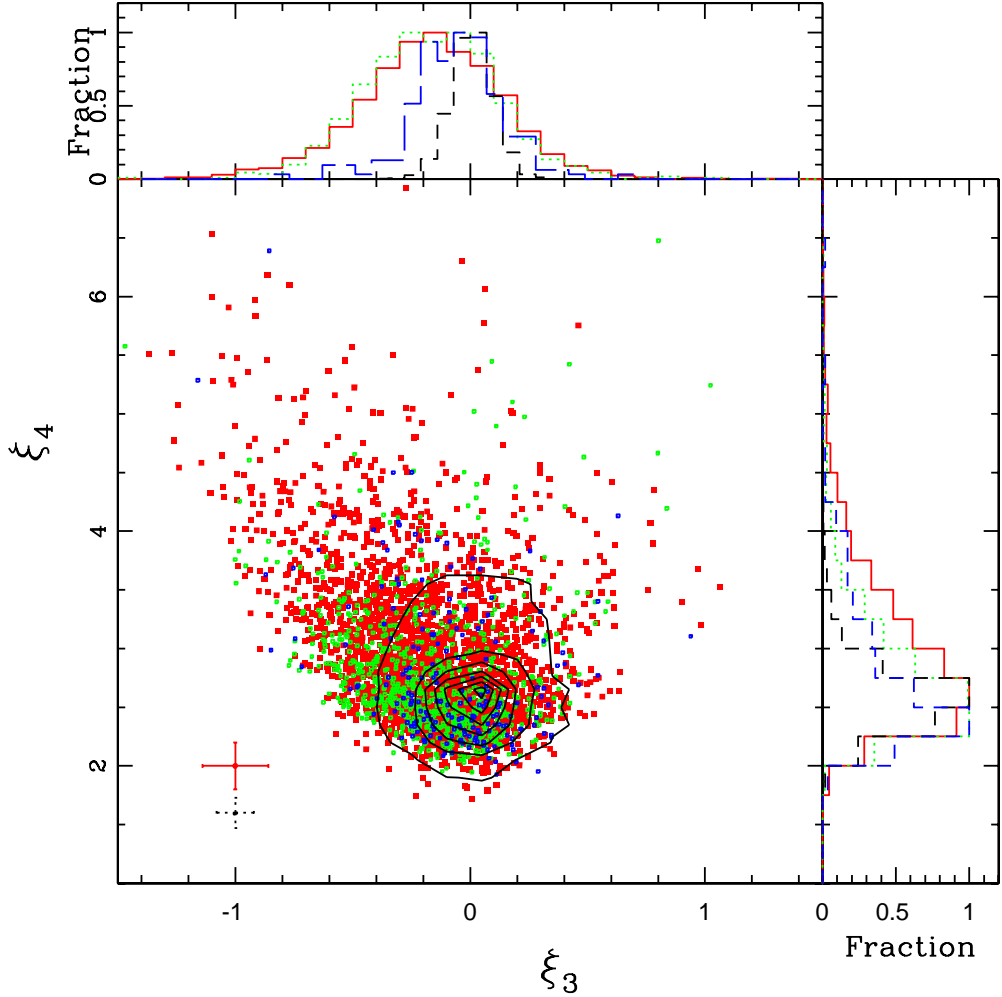}
\caption{\it Bottom-left panel:\rm\ $\xi_4$ versus $\xi_3$ diagram for all the sub-samples. The AGNs and transition galaxies 
are presented by the red and green points, respectively. The blue points are plotted for the star-forming galaxies with 
[\ion{O}{3}] line widths $\sigma>2\sigma_{\mathrm{inst}}$. 
The over-plotted black contours show the distribution for the star-forming galaxies in the comparison sample
with $\sigma<2\sigma_{\mathrm{inst}}$. 
The red-solid cross and the black-dashed cross at the left-bottom corner indicates the uncertainties for the large line-width samples and 
the comparison sample, respectively. The uncertainties are estimated from the duplicates. \it Upper-left panel:\rm\
distributions of the parameter $\xi_3$ for the four sub-samples (AGNs: red solid line, transition galaxies: 
green dotted line, star-forming galaxies with $\sigma>2\sigma_{\mathrm{inst}}$: blue long-dashed line, and 
star-forming galaxies with $\sigma<2\sigma_{\mathrm{inst}}$: black short-dashed line). \it Bottom-right panel:\rm \
The same as the upper-left panel but for the parameter $\xi_4$.
}
\end{figure}

\begin{figure}
\epsscale{.80}
\plotone{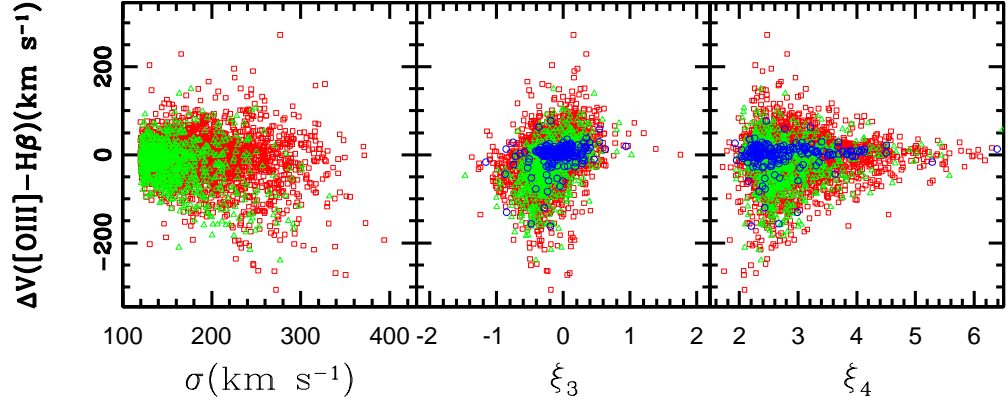}
\caption{Relative velocity shift $\Delta\upsilon$ plotted against line width $\sigma$ (\it left panel\rm), parameters
$\xi_3$ (\it middle panel\rm) and $\xi_4$ (\it right panel\rm). In each panel, the AGNs and 
transition galaxies are plotted by the red and green points, respectively. The star-forming galaxies with 
$\sigma\geq2\sigma_{\mathrm{inst}}$ are marked by the blue points.  
}
\end{figure}

\begin{figure}
\epsscale{.80}
\plotone{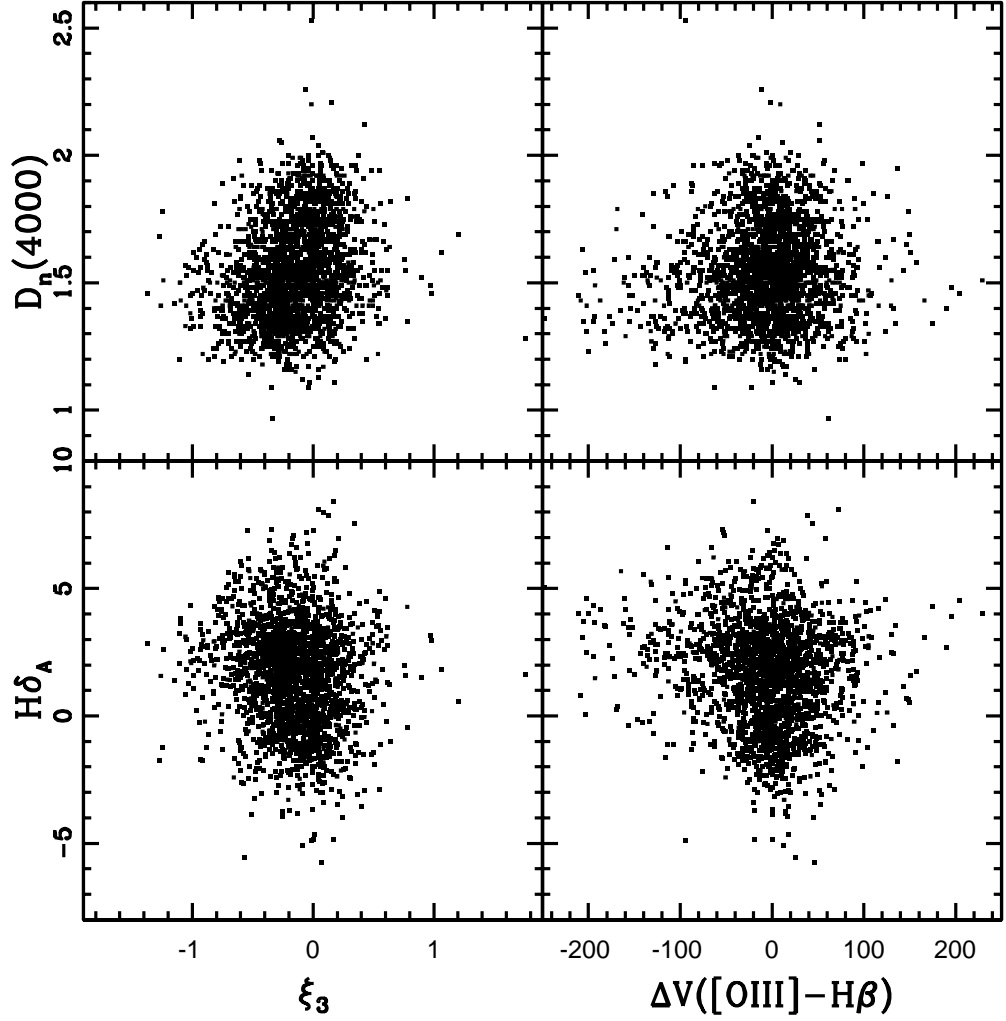}
\caption{\it Left column:\rm\ stellar population properties (i.e., the indices of $D_n(4000)$ and 
H$\delta_\mathrm{A}$) plotted
against the line shape parameter $\xi_3$ for the AGNs. 
\it Right column:\rm\ both indices plotted against the relative velocity shift of 
the [\ion{O}{3}] emission line for the AGNs.
}
\end{figure}

\begin{figure}
\epsscale{.80}
\plotone{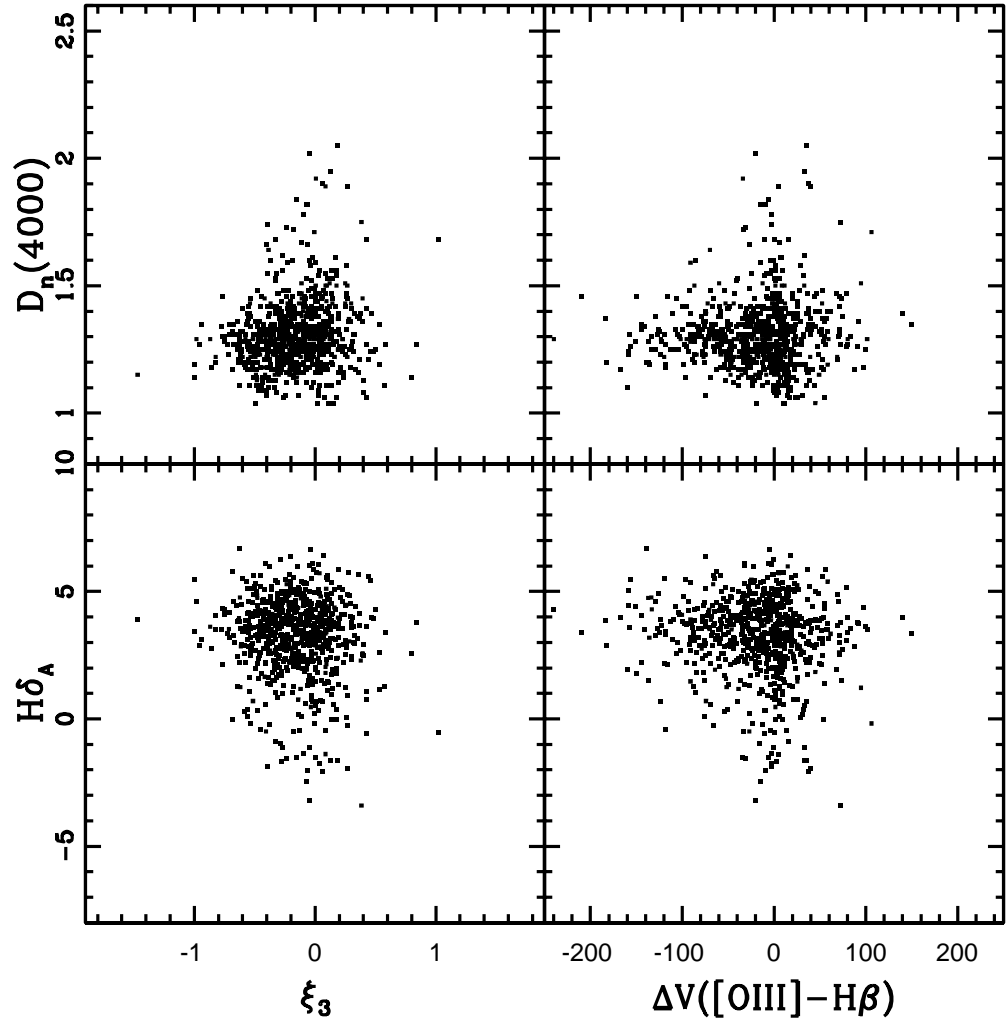}
\caption{The same diagram as Figure 5 but for the transition galaxies. 
}
\end{figure}

\begin{figure}
\epsscale{.80}
\plotone{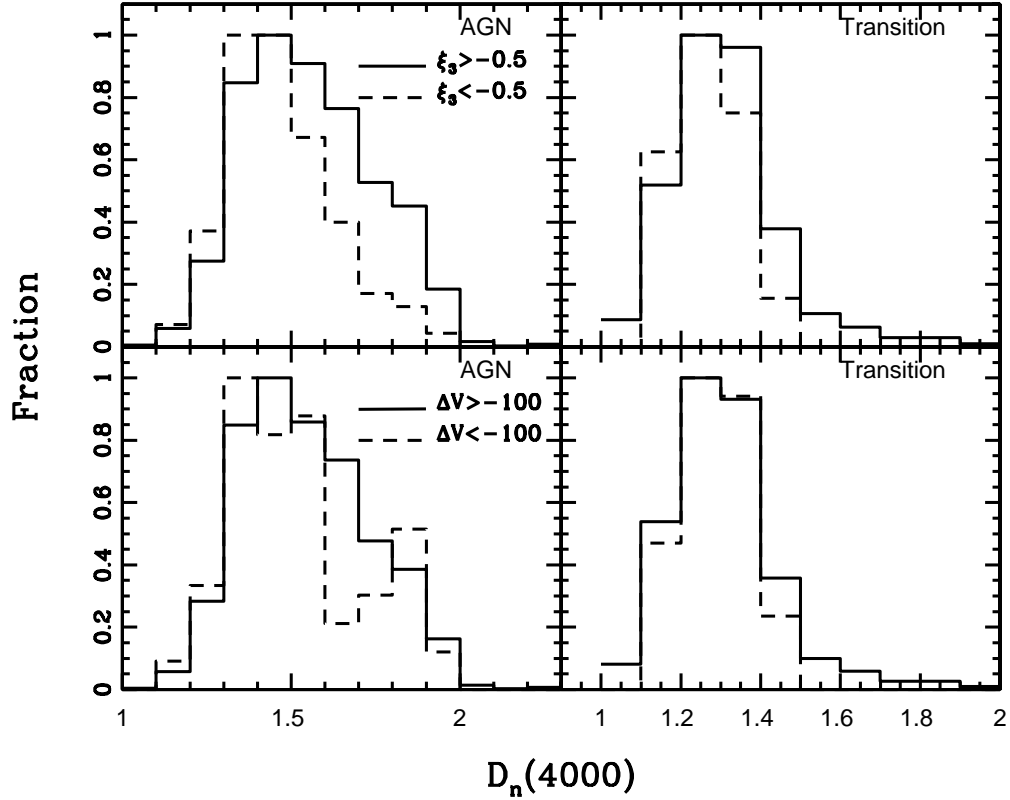}
\caption{The distributions of $D_n(4000)$ are compared between the two groups with different [\ion{O}{3}]
line shape parameters (i.e., $\xi_3$ for the upper raw), and between the two groups with 
different relative velocity shifts (i.e., $\Delta V$([\ion{O}{3}]-H$\beta$) 
for the bottom raw) for the AGNs (left column) and the transition galaxies (right column). In each panel,
the solid line represents 
the objects with $\xi_3>-0.5$ (the upper raw) or $\Delta V> -100\ \mathrm{km\ s^{-1}}$ (the bottom raw), and
the dashed one the objects with $\xi_3<-0.5$ or $\Delta V< -100\ \mathrm{km\ s^{-1}}$. 
}
\end{figure}

\begin{figure}
\epsscale{.80}
\plotone{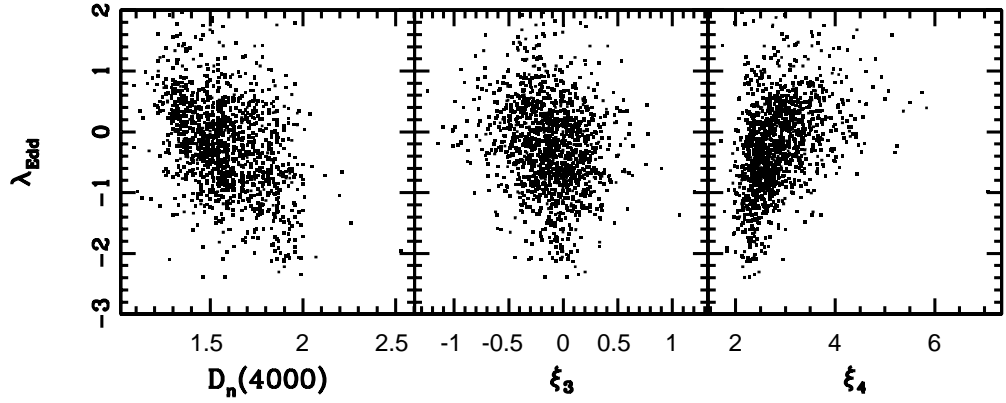}
\caption{$L_{\mathrm{[OIII]}}/\sigma_*^4$ is plotted as a function of $D_n(4000)$ (\it left panel\rm), the line profile
parameters $\xi_3$ (\it middle panel\rm) and $\xi_4$ (\it right panel\rm) for the AGN sample. }
\end{figure}

\clearpage

\begin{table}
\begin{center}
\small
\caption{Statistical properties of the shape parameter $\xi_3$ for all the sub-samples.\label{tbl-1}}
\begin{tabular}{ccccc}
\tableline\tableline
Property & AGNs & Transitions & SF galaxies/$>2.0\sigma$ & SF galaxies/$<2.0\sigma$\\
(1) & (2) & (3) & (4) & (5)\\
\tableline
Numbers & 2333 & 793 & 190 & 5843\\
$\overline{\xi}_3$ & -0.167 & -0.170 & -0.09 & 0.01\\
$\langle\xi_3\rangle$ & -0.165 & -0.167 & -0.08 & 0.01\\
$\sigma_{\xi_3}$ & 0.31 & 0.28 & 0.32 & 0.14 \\
\tableline
\end{tabular}
\end{center}
\end{table}

\begin{table}
\begin{center}
\small
\caption{Statistical properties of the shape parameter $\xi_4$ for all the sub-samples.\label{tbl-1}}
\begin{tabular}{ccccc}
\tableline\tableline
Property & AGNs & Transitions & SF galaxies/$>2.0\sigma$ & SF galaxies/$<2.0\sigma$\\
(1) & (2) & (3) & (4) & (5)\\
\tableline
Numbers & 2333 & 793 & 190 & 5843\\
$\overline{\xi}_4$ & 3.00 & 2.79 & 2.84 & 2.61\\
$\langle\xi_4\rangle$ & 2.83 & 2.64 & 2.56 & 2.58\\
$\sigma_{\xi_4}$ & 0.70 & 0.56 & 1.01 & 0.43 \\
\tableline
\end{tabular}
\end{center}
\end{table}

\begin{table}
\begin{center}
\small
\caption{Two-sides Kolmogorov-Smirnov test matrix for the shape parameter $\xi_3$.\label{tbl-1}}
\begin{tabular}{cccc}
\tableline\tableline
Sub-sample & AGNs & Transitions & SF galaxies/$>2.0\sigma$\\
 &(1) & (2) & (3) \\
\tableline
AGNs & - & 0.034(0.48) & 0.159($2.9\times10^{-4}$)\\
Transitions & - & - & 0.164($5.0\times10^{-4}$)\\
SF galaxies/$>2.0\sigma$ & - & - & -\\
\tableline
\end{tabular}
\end{center}
\end{table}

\begin{table}
\begin{center}
\small
\caption{Two-sides Kolmogorov-Smirnov test matrix for the shape parameter $\xi_4$.\label{tbl-1}}
\begin{tabular}{cccc}
\tableline\tableline
Sub-sample & AGNs & Transitions & SF galaxies/$>2.0\sigma$\\
& (1) & (2) & (3) \\
\tableline
AGNs & - & 0.154($<10^{-9}$) & 0.214($2.0\times10^{-7}$)\\
Transitions & - & - & 0.133($9.0\times10^{-3}$)\\
SF galaxies/$>2.0\sigma$ & - & - & -\\
\tableline
\end{tabular}
\end{center}
\end{table}

\begin{table}
\begin{center}
\small
\caption{Spearman Rank-order Correlation Coefficient Matrix for the AGNs\label{tbl-1}}
\begin{tabular}{ccc}
\tableline\tableline
  Property & $D_{4000}$ & $\lambda_{\mathrm{Edd}}$ \\
 & (1) & (2) \\
\tableline
$\xi_3$ & 0.218 & -0.204 \\
$\xi_4$ & -0.292 & 0.449\\
\tableline
\end{tabular}
\end{center}
\end{table}

\clearpage







\end{document}